\documentclass{aa}
\usepackage{graphics}

\begin{document}

\title{Position-sensitive detector for the 6-meter optical telescope }

\author{V. Debur\inst{1} \and T. Arkhipova\inst{2} \and G. Beskin\inst{1, 4} \and M. Pakhomov\inst{2} \and V. Plokhotnichenko\inst{1, 4}
 \and M. Smirnova\inst{2} \and A. Solin\inst{3}}

\institute{Special astrophysical observatory, Nizhnij Arkhyz, 369167, Russia; 
\and Scientific
research institute of electron devices, Moscow, 111123, Russia; 
\and National centre
of high energy physics, Minsk, 220040, Belarus; 
\and Isaac Newton Institute of
Chile, SAO Branch, Russia.}

\date{}

\abstract{
The Position-Sensitive Detector (PSD) for photometrical and spectral
observation on the 6-meter optical telescope of the Special Astrophysical Observatory
(Russia) is described. The PSD consists of a position-sensitive tube, amplifiers
of output signals, analog-to-digital converters (ADC) and a digital logic plate,
which produces a signal for ADC start and an external strob pulse for reading
information by registration system. If necessary, the thermoelectric cooler
can be used. The position-sensitive tube has the following main elements: a
photocathode, electrodes of inverting optics, a block of microchannel plates
(MCP) and a position-sensitive collector of quadrant type. The main parameters
of the PSD are the diameter of the sensitive surface is 25 mm, the spatial
resolution is better than 100 \( \mu  \)m in the centre and a little worse
on the periphery; the dead time is near 0.5 \( \mu  \)s; the detection quantum
efficiency is defined by the photocathode and it is not less than 0.1, as a
rule; dark current is about hundreds of cps, or less, when cooling. PSD spectral
sensitivity depends on the type of photocathode and input window material. We
use a multialkali photocathode and a fiber or UV-glass, which gives the short-
wave cut of 360 nm or 250 nm, respectively. 
}

\maketitle

\section{Introduction. }

The light fluxes from the celestial objects under study are so weak that even
in observations with large telescopes a photon falls on one element of the image
far from every second. For such fluxes a maximum quantum efficiency is an obvious
requirement to the detector. And in our investigations of relativistic objects
for the purpose of search for variability of quantum fluxes up to microseconds
it is necessary to analyse the time series of quantum registration moments.
In this case the variability manifests itself in the short-wave region of the
optical spectrum. These requirements defined use of position-sensi- tive detectors
(PSD) in observations with our 6-m op- tical telescope. The PSDs represent vacuum
photoelectron tubes with microchannel amplification and position-sensitive collectors.
The advantage of this type of detectors is that the sensitivity of their photocathodes
is sufficiently high, and the most important is the fact that there is a possibility
of detection of quantum arrival times with an accuracy up to dozens of picoseconds
in prospect. This fact makes the PSD suitable for investigations of relativistic
objects and we are working in this field now.

\section{Design of the PSD and duties of its main components. }

\begin{figure}
{\centering \resizebox*{!}{0.5\textheight}{\rotatebox{270}{\includegraphics{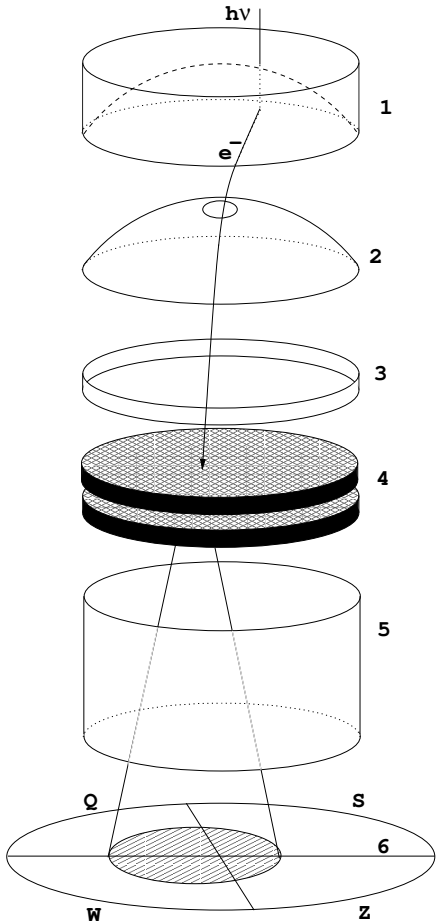}}} \par}
\caption{\label{fig: PSD_SH}Shematic view position sensetive detector.}
\end{figure}

A schematic view of the detector is given in Fig.\ref{fig: PSD_SH}. 

The numbers designate the following: 1 - input optical disk with a photocathode
on its internal surface; 2 - anode (by changing its voltage one can change the
electro-optical magnification within 0.5-1.5); 3 - antidistortion ring; 4 -
stack of microchannel amplification formed by three or two microchannel plates
(MCP); 5 - metal rings which create an electrostatic field of such a configuration
which provides a maximum possible working field at minimum geometrical distortions
and charge losses at the periphery; 6 - position-sensitive collector of quadrant
type. This collector has an advantage over a wedge and strip ones {[}T. Arkhipova
et al.{]} in considerably smaller interelectrode capacitances. This permits one
to diminish the amplification of the MCP-stack and extend the range of operation
of the detector towards greater load. A lower capacitance will lead to decrease
of noises of the amplifiers, which gives an improvement of spatial resolution.
The disadvantage of the quadrant collector is the deterioration of the resolution
at the periphery; however it is of minor importance when observing star-like
objects because the object observed can be placed in the centre.

\section{The electronic schemes and design of the photosensitive device. }

\begin{figure}
{\centering \resizebox*{0.9\columnwidth}{!}{\includegraphics{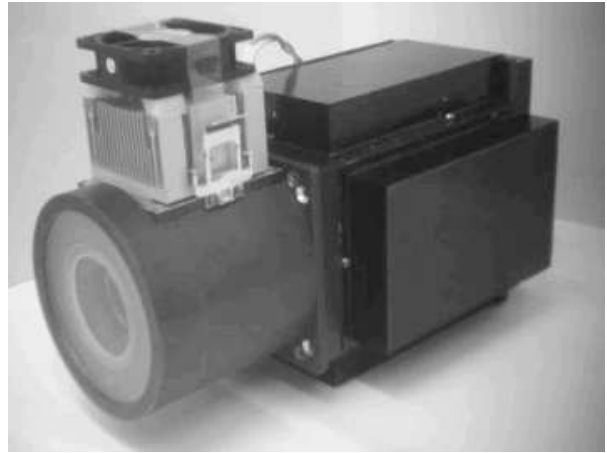}} \par}

\caption{\label{fig: PSD_view}General view Position sensetivity detector.}
\end{figure}

The important components of the device are signal amplifiers of the PSD collector.
We use integrated curcuit amplifiers of The National Centre of High Energy Physics
(NCHEP) {[}O. Dvornikov et al{]}. A computer modeling with allowance made for
the PSD parameters having a quadrant collector made it possible to choose an
optimum value of the time of shaping of 0.25\( \mu  \)s, an equivalent noise
of no more than 600 electrons was provided. With an amplification of the MCP
stack by several units 106, the statistical noise for the central area of the
collector is equal approximately to the same value and this allows an image
format no smaller than 300x300 to be obtained. Such a resolution complies with
the astrophysical requirements. The amplifier is made like charge-sensitive
ones (CSA) and consists of two sections {[}V. Debur, A. Solin{]}. The first
section is intended for the transformation of the input charge to the output
voltage with a decay time of 2.5\( \mu  \)s. When passing to the second section
the pulse is shortened to a value of 0.25\( \mu  \)s with quasi-Gaussian shaping.
The amplifier provides a transformation coefficient of 30 mV/fC necessary for
operation with the analog to digital converters (ADCs) having an input signal
range of 2 V, with the nominal gain of the MCP stack. The amplifiers are not
overloaded and make it possible to register adjacent pulses separated by the
time a few times as short as the pulse duration. For digitization of signal
we use 10-digit ADCs of type AD871. The equivalent noise charge of the CSA equal
to 600 electrons corresponds to a value of about 3 mV at 30mV/fC, that is the
ADCs provide the measurements of signals with sufficient accuracy. The check
of the registration channels by way of supplying test signals to the CSA inputs
yielded a FWHM value of about 2-3 channels for the each of four amplifiers.
The timing channel of the detector is needed for the formation of the start
pulses of the ADCs and production of strobe pulses for recording the time ~moments
of the registered quants. The amplifier and discriminator connected to the output
plate of the MCP stack are also designed by using of ICs of the NCHEP. The whole
description of electronics questions is given in {[}V. Debur, A. Solin {]}.
The external view of the device is displayed in Fig. \ref{fig: PSD_view}. The entrance window covers the space around the PSD, which is cooled by a single-stage
thermoelectrical refrigerator. The heat from the hot surface is withdrawn with
the help of the processor cooler. The housing of the PSD contains the PSD and
amplifiers. The side compartments hold the high voltage supply resistors of
the PSD, the ADCs and auxiliary plates. At the back are located plug-and- sockets
for connection of power supply and output signals to the CAMAC-crate in which
the modules of the system Quantochron and necessary service modules are placed.
The overall dimensions of the device are 140x180x290 (mm). 

In development of the photosensitive device it was required that it should be
operated with the available registration system Quantochron {[}A. Zhuravkov
et al{]}. The output information must represent 16-digits words with TTL-levels
and strobe-pulse for measuring the time moments.

\section{Data registration}

\begin{figure}
{\centering \resizebox*{0.9\columnwidth}{!}{\includegraphics{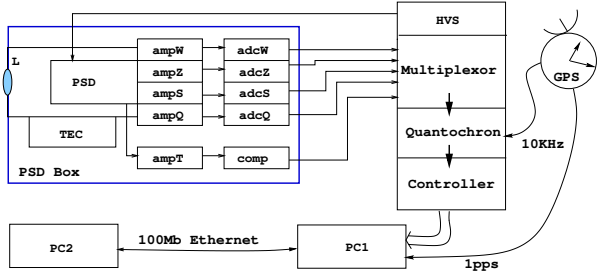}} \par}

\caption{\label{fig: compl}The functional scheme of aquisition complex.}
\end{figure}

The PSD operates as part of the acquisition complex incorpora ting also a device
for receiving photocount flux codes, a computer for control and data acquisition,
which is located in the local net with the computer of the astronomer- operator.
The functional diagram of the photometrical complex is shown in Fig. \ref{fig: compl}. The flux of photocounts from the PSD must be transferred to the computer data
storage in the form it is received by the detector, as counts of registrat ion
time of all the quanta and their coordinates, the rate of count arriving must
be up to 100 000 quants/s. For this purpose, we use the time-code converter
Quantochron {[}1{]}. Primarily it was designed for registration of the data
flows which have a 16-bit coordinate field (265x256 elements). The application
of the PSD with a quadrant collector and the use of 10-bits ADC demanded extension
of capacity up to 40 bits, this is why as a temporal measure we use a multiplexor
which disconnects each arriving 40-bits photocount into three sequential messages:
8, 16 and 16 bits. The 8 bits in the first message are used for auxiliary information.

\section{Characteristics of the PSD. }

\begin{figure}
{\centering \resizebox*{0.9\columnwidth}{!}{\includegraphics{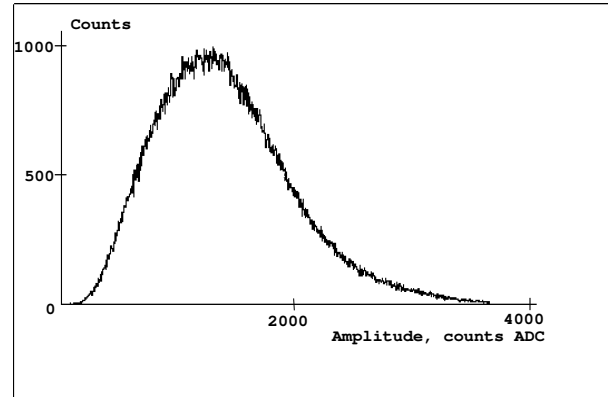}} \par}

\caption{\label{fig: charge}The charge distribution of phtocounts.}
\end{figure}

\begin{figure}
{\centering \resizebox*{0.9\columnwidth}{!}{\includegraphics{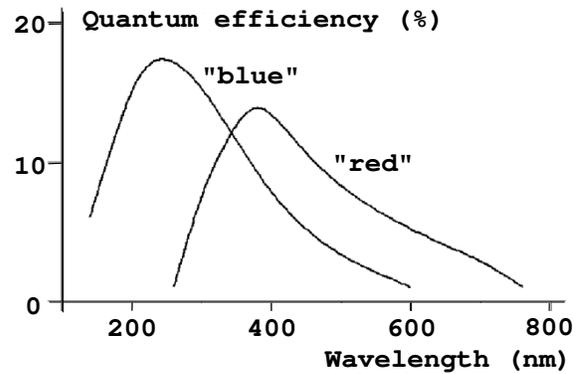}} \par}

\caption{\label{fig: Spectral}Specrtal sensetivity of PSD.}
\end{figure}

The single electron pulses distribution of the detector has been derived by
summation of all 4 channels and is presented in Fig. \ref{fig: charge}. Despite
the fact that all individual channels have a nearly exponential distribution,
the common distribution has a well defined peak, which allows a conclusion to
be made that the acquisition system registers practically all the pulses. Nevertheless
the branch from the side of small amplitudes impairs the spatial resolution
at the periphery. The shape of the amplitude distribution is preserved at count
rates up to 100 cps and above, the photocathode being uniformly illuminated.
With point-like illumination, like the star image, at count rates higher than
10 cps the images of stars begin to differ from gaussian by a flatter peak and
then begin to acquire a dip at the centre. The characteristics of the spectral response of the photocathodes of two different
types are presented in Fig. \ref{fig: Spectral}. The photocathode of the \char`\"{}red\char`\"{} PSD is on a fiber disc of ordinary
glass, while the photocathode of the \char`\"{}blue\char`\"{} PSD is on a disc
of the UV glass. The latter was made to achieve maximum sensitivity in the ultraviolet
region. The presented types of the photocathodes largely differ in thermoelectron
emission current. One can work with the \char`\"{}blue\char`\"{} photocathode
at a room temperature without cooling. It has about 50 electrons/s from the
whole area, whereas the \char`\"{}red\char`\"{} photocathode has about 15000
electrons/s at a room temperature and it is necessary cooling down to \( 0^{0} \)C. 

\begin{figure}
{\centering \resizebox*{0.9\columnwidth}{!}{\includegraphics{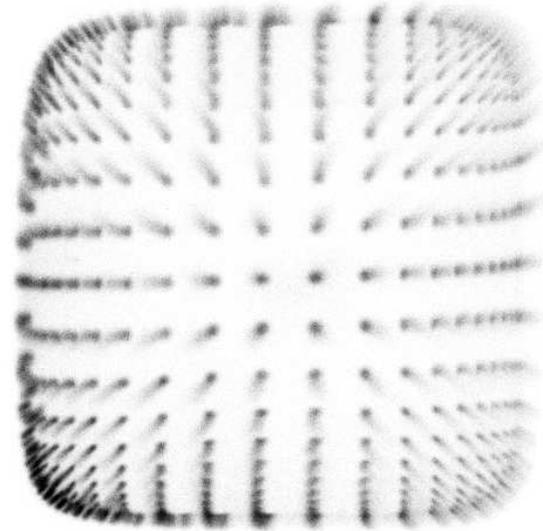}} \par}
\caption{\label{fig: reper_no}The reper field before any correctoins}
\end{figure}

Figure \ref{fig: reper_no} shows 512x512 elements detector field with image of
emitting points got as a result of photons coordinates binning.

Reduction of collected data is performed for each received photon separately
and is based on knowledge of counts distortion mechanisms related to 
electronic clound formation geometry.

Some of this process features are:

\begin{itemize}
\item Different scales of image for different integral charge levels of each photo count -
for lower charges image scale is bigger than for higher ones. Functional shape
of inverse charge value and point coordinates for one physical source is
nearly linear. Such scale dependency correction makes it possible to
increase image sharpness over the whole detector field.

\item Spherical image distortion may be successfully compensated by tangential correction.

\item Cross-like central condensation on flat-field image is due to superposition of
electronic density gap in electronic clouds centres and dielectric gap between
detector quadrants and may be corrected using arctangent coordinates transformation.
This transformation must also take into account non-perpendicularity of collector
quadrants edges.

\end{itemize}

\begin{figure}
{\centering \resizebox*{0.9\columnwidth}{!}{\includegraphics{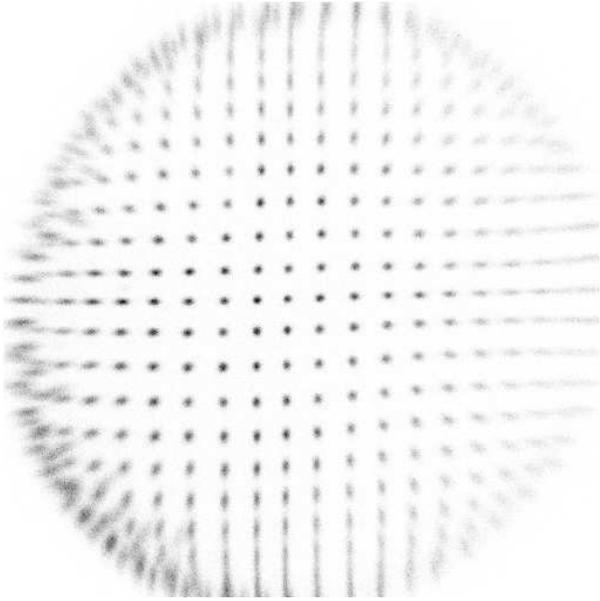}} \par}
\caption{\label{fig: reper_corr}The reper field after correction. Mask pupils size
is 0.1 mm, distande betwee pupils - 1 mm.}
\end{figure}

Fig. \ref{fig: reper_corr} shows the result of applying these transformations
to image.

Residual non-singular central condensation may be compensated by polynomial
corrction.

The FWHM of the point spread function at the centre is less than 100 \( \mu  \)m.
Such quality is quite satisfactory since the size of the stars in our optical
system is generally 400-500 \( \mu  \)m, and only on rare nights it may diminish
to 150 \( \mu  \)m. The field that we observe usually contains one investigated
star. At the present time we are working over the increase of information capacity
of images being acquired. For this purpose, we use polarization and light-dividing
elements which allow observation of the object and comparison star simultaneously
in one field, in different polarization and spectral intervals. In so doing,
the number of object images increases to eight {[}V. Plokhotnichenko et al{]}.

\section{Prospects of development of the PSD at SAO. }

The improvement of the spatial resolution can be attained through increasing
the gain of the MCP stack and may come quite closely to the value defined by
the size of the MCP channels, about 15 \( \mu  \)m and less. The requirements
to the noise of the charge sensitive amplifiers become more stringent. The improvement
of the temporal resolution can be brought to less than nanosecond if required.
However, in our investigations it makes no sense to make it an order of magnitude
better than the dead time of the system. And we evaluate the reserve of improvement
of the dead time as nearly an order, since by the present time integral circuits
of multichannel amplifiers capable of operating at shaping times of about 100
ns have been developed. The increase in the sensitive area diameter to 40 mm
seems to be relatively simple, but the increase up to 80 mm is possible only
with the use of electro-optical diminishing in the input camera. The quantum
efficiency depends on the quality of the photocathode, and its possible to have
almost 40\% at maximum. The dark current must be much lower than the night sky
background, but the cooling of the photocathode is a relatively simple task.
The increase in the upper count rate for a point image presents a problem, however,
it is possible by nearly an order of magnitude for spread images provided that
the dead time is decreased. It appears worth-while for the authors to pass to
a multielement collector, for instance, 8-elements hexagonal or up to 19-elements.
The progress achieved in integral circuits and computers makes possible registration
and reduction of signals from such collectors. 

\section*{Acknowledgements}
This investigation was supported by the Russian Ministry of Science, Russian
Foundation of Basic Researches (grant 01-02-17857), Federal Program \char`\"{}Astronomy\char`\"{}
and INTAS (grant 96-542).

\end{document}